\documentclass{aa}
\usepackage{graphicx}
\begin{document}
   \title{An
accretion disk model for periodic timing variations of pulsars}

    \author{Qiao G.J., Xue Y.Q., Xu R.X., Wang H.G. \& Xiao B.W.
             }

  \offprints{Qiao G.J.}

   \institute{Department of Astronomy, School of Physics, Peking
University, Beijing 100871, China\\
              \email{gjn@pku.edu.cn, bxyq@vega.bac.pku.edu.cn}
                 }

   \date{Received;accepted}

   \abstract{
   The long$-$term, highly periodic and correlated variations in both
the pulse shape and the rate of slow$-$down of two isolated
pulsars (PSRs) PSR B1828$-$11 and PSR B1642$-$03 were discovered
recently. This phenomenon may provide evidence for ``free
precession'' as suggested in the literature.
%
%But this is difficult to understand theoretically, because
%torque-free precession of a solitary pulsar should be damped out
%by the vortices in its super-fluid interior (Stairs et al. 2000).
%
Some authors presented various kinds of models to explain this
phenomenon within the framework of free precession. Here we
present an accretion disk model for this precession phenomenon
instead. Under reasonable parameters, the observed phenomenon can
be explained by an isolated pulsar with a fossil disk. This may
link radio pulsars and anomalous X$-$ray pulsars (AXPs) and
present an indirect evidence for the existence of the fossil disk
in nature.
   }

   \maketitle
%
%________________________________________________________________

\section{Introduction}

Two isolated pulsars (PSR B1828$-$11 \& B1642$-$03) present the
phenomenon of long$-$term, highly periodic and correlated
variations in both the pulse shape and the slow$-$down rate (for
PSR B1828$-$11, its variations are best described as harmonically
related sinusoids, with periods of approximately 1,000, 500 and
250 days, see Stairs et al. 2000; for PSR B1642$-$03, its timing
residuals exhibit cyclical changes with amplitude varying from 15
to 80 ms and spacing of maxima varying from 3 to 7 yr, see Cordes
1993 and Shabanova et al. 2001). This phenomenon provides the most
compelling evidence for precession (Link \& Epstein 2001). Some
authors (Jones \& Andersson 2001, Link \& Epstein 2001, Rezania
2002) proposed different models to explain this phenomenon within
the framework of free precession.
%
%Unfortunately, this scenario seriously challenges our
%understanding of the liquid interior of the neutron star.
%Torque-free precession of a solitary pulsar should be damped out
%due to the dissipation caused by the vortices pinning to the
%stellar crust on the timescale of several hundred precession
%periods (Stairs et al. 2000).
%
However, further investigations are expected on whether the force
of vortices pinning to the stellar crust is strong enough to damp
out the free precession of neutron stars (see the conclusion
section for details).

The pulsating X$-$ray binary Her X$-$1 also shows an unusual kind
of periodic timing behavior in the form of an on$-$off cycle about
35 days. Sarazin et al. (1980) showed that a moderately massive
disk ($10^{-5}-10^{-2}$ $M_{\odot}$) can cause the precession of
the central compact star. Trumper et al. (1983) used the free
precession of the neutron star to explain the 35$-$day period.
Qiao \& Cheng (1989) suggested that instead of the free precession
of the neutron star, the periodic variation is due to the
precession of the accretion disk driven by a normal star.

In this paper, we present an accretion disk model to account for
the precession of pulsars. The total torque exerted on the pulsar
by the fossil disk, which causes the pulsar to precess, is
nonzero, due to the asymmetry in pulsar shape and nonalignment
between the disk rotation axis and the axis of pulsar moment of
inertia. Hence, we present this forced precession model
(disk$-$induced precession), other than the free precession model.

In fact, such a system, i.e. a pulsar with an accretion disk, has
been invoked to describe the nature of two kinds of isolated
X$-$ray pulsars: the anomalous X$-$ray pulsars (AXPs, which have
persistent X$-$ray pulsations but few burst) and the soft
gamma$-$ray repeaters (SGRs, which have multiple bursts of soft
gamma$-$rays) (see Xue et al. 2003 for a review). The most
remarkable difference between these two kinds of radio$-$quiet
X$-$ray pulsars and the radio pulsars is that their X$-$ray
luminosities (usually $L_{\rm X}$ $\sim$ 10$^{35}$ ergs/s) are
several orders of magnitude higher than their rotational energy
loss rates. To explain this and other exotic facts, two
alternative models, viz. the magnetar models (in which the pulsar
is suggested to be a neutron star with super$-$strong magnetic
field) and the accretion disk models, were proposed (see Xue et
al. 2003 for a review). Under the framework of the accretion disk
model, a problem would come if the precession of radio pulsars
(e.g. PSR B1811$-$11 and B1642$-$03) was really caused by the
accretion disk: why do some pulsars with disks behave as ordinary
radio pulsars while others behave as AXPs/SGRs? In this paper, we
discuss this problem in section 3 after presenting the detailed
calculation in section 2.

%__________________________________________________________________

\section{Pulsar precession caused by an accretion disk}

\subsection{Geometry and parameters}

Consider the neutron star as a rotation ellipsoid with the
principal moment of inertia $I_{1}=I_{2} \neq I_{3}$ and the
corresponding radii $a=a \neq b$. The geometry of the system is
shown in Fig.1: $O$$-$$xyz$ is the coordinate system fixed with
the neutron star, $Ox$, $Oy$ and $Oz$ are the principal axes of
moment of inertia of the neutron star, the axis $OA$ is the
rotation axis of the neutron star (for simplicity, we assume the
axis $Oz$ is aligned with the axis $OA$ in this paper; in this
case, free precession doesn't exist and we can see how the disk
makes the neutron star precess more clearly), $O$$-$$\xi\eta\zeta$
is the coordinate system fixed with the disk and the disk plane is
$O\xi\eta$, $\theta$, $\varphi$ and $\phi$ are Euler angles,
$\omega$ is the angular velocity of neutron star rotation. We will
derive the formulae of the gravitational potential in the neutron
star$-$disk system and the angular velocity of neutron star
precession. And then, as an example, we model the precession
period of PSR B1818$-$11. The parameters about PSR B1828$-$11 are
as follows: the rotational period of this pulsar $P=0.405$s
($\omega=2\pi/P$), the period derivative
$\dot{P}=6.0\times10^{-14}$ ss$^{-1}$, the precession period
$P_{\rm pre} \approx 1000$ days, the magnetic field
$B=5.0\times10^{12}$G (Stairs et al. 2000).

%%%%%FIG1 SHOULD BE PUT HERE! ******************************FIG1 SHOULD BE PUT HERE!**
%______________________________________________

\subsection{Equations and numerical results}

   The gravitational potential between the neutron star and the disk is

\begin{eqnarray}
V&=&-\frac{2GMM_{0}}{c+d}\nonumber\\
&\times&[1-\frac{b^{2}+2a^{2}}{10cd}+\frac{3b^{2}}
{20cd}\sin^{2}\theta+\frac{3a^{2}}{20cd}(1+\cos^{2}\theta)],
\end{eqnarray}
where $M$, $a$ and $b$ are the mass, the equatorial and
longitudinal radii of the neutron star, respectively, $M_{0}$ is
the mass of the disk, $c$ and $d$ are the inner and outer radii of
the disk, respectively.

The moment of inertia of the neutron star is $I_{\rm z}=(8/15)\pi
\rho a^{4} b$, where $\rho=3M/(4 \pi a^{2} b)$ is the density of
the neutron star. Since the rotation angular velocity of the
neutron star $\omega$ is much greater than the precession angular
velocity of the neutron star $\dot{\phi}$, the angular momentum of
the neutron star can be written as $L_{0}\simeq I_{\rm z}\omega$.
According to the Lagrange's equation
$L_{0}\dot{\phi}\sin\theta=-\partial V/
\partial \theta$ (Qiao \& Cheng 1989), we get the precession
angular velocity of the neutron star:

\begin{equation}
\dot{\phi}=-\frac{3GM_{0}\cos\theta}{2cd(d+c)\omega}(1-\frac{b^{2}}{a^{2}}).
\end{equation}

In terms of the study on the stability of the Maclaurin spheroids,
we can obtain the oblateness of this pulsar:

\begin{equation}
\epsilon=\frac{a-b}{a}=2.27\times10^{-6}.
\end{equation}

For simplification, we assume $c\approx d\approx R$, so that

\begin{equation}
\dot{\phi}=-\frac{3GM_{0}\cos\theta}{4R^{3}\omega}[1-(1-\epsilon)^{2}].
\end{equation}

Fig.2 shows the relations between $M_{0}$ and $R$ derived from the
1000$-$day precession period for a group of $\theta$. In
principle, $M_{0}$ may vary from less than $0.001M_{\odot}$ to
even $0.1M_{\odot}$ (The amount of the fall$-$back material
($M_{\rm fb}$) which forms the disk after supernova explosion is
uncertain, however, $M_{\rm fb}\leq 0.1M_{\odot}$ is supposed to
be reasonable, see Chatterjee et al. 2000), and $R$ from about
$1R_{\rm co}$ (or smaller) to $2\sim 3R_{\rm co}$ when $\theta$
varies within the range of $0\sim 90^{o}$, where $R_{\rm co}$ is
the corotation radius of the disk with the pulsar. Thus, the
precession period of the PSR B1828$-$11 is well elucidated.

%%%%%FIG2 SHOULD BE PUT HERE! ******************FIG2 SHOULD BE PUT HERE!**

\section{Possible evolutionary scenario in the pulsar$-$disk system}

One may ask where the accretion disk comes from. Usually, a disk
around the pulsar can be formed: 1. by accreting the matter from a
companion star in the binary system (approach I); 2. by the
fall$-$back of the material after a supernova explosion (approach
II). In the former process, there are two ways of mass transfer:
matter flowing through the first Lagrange point due to the
critical Roche Lobe abounding with the companion matter, or by the
stellar wind (Nagase 1989).

An interesting issue is that what the pulsar in the pulsar$-$disk
system will behave as, radio pulsar with precession or AXP/SGR? In
the accretion model of X$-$ray binary, accreting process may
experience some ``phases'', e.g. ``radio pulsar phase'',
``sleeping phase'' , ``propeller phase'' and ``accretion phase''
(Illarionov \& Sunyaev 1975, Henriches 1983). The case is
suggested to be similar for a pulsar$-$disk system formed through
either approach I or II. The PSR B1828$-$11, B1642$-$03, AXPs and
SGRs are all isolated ``neutron stars''. What are the links
between them? Here we present a possible series of evolutionary
tracks for the pulsar$-$disk system (cf. Chatterjee et al. 2000
and Henriches 1983).

(a) After the supernova explosion, the newborn neutron star
rotates so fast that its magnetospheric radius $R_{\rm m}$ is much
larger than its light cylinder radius $R_{\rm LC}$. The coupling
between the disk and the neutron star is unlikely. The young
pulsar initially emits a very strong relativistic wind together
with highly energetic magnetic dipole radiation. The pulsar wind
and photons blow a cavity in the fall$-$back flow and no matter
can accrete, as is similar in the binary system (Henriches 1983).
Thus the neutron star may live in the ``radio pulsar phase'', i.e.
the radio emission can propagate away.

(b) The neutron star slows down as its rotational energy reduces.
Its radiation and the high energy particle flow become weakened,
thus the cavity decreases and the radio emission is finally
quenched. When the ``radio pulsar phase'' ends, accretion down to
the surface of the neutron star is not expected immediately. The
so$-$called ``sleeping phase'' takes place. This phase lasts until
the pulsar rotation slows down enough to make accretion possible.

(c) Subsequently, the neutron star enters the ``propeller phase''
since accretion onto the neutron star is inefficient due to
centrifugal forces acting on the matter. As the rotation slows
down further, accretion will be permitted and the neutron star
will behave as an accretion$-$driven X$-$ray source. Gradually,
the system enters a quasi$-$equilibrium ``tracking phase''
(happening at the time of $t_{\rm trans}$), in which the spin of
the star roughly matches the rotation of the disk at $R_{\rm m}$.
In this phase, the neutron star will be X$-$ray bright due to
accretion. It is possible that the advection$-$dominated accretion
flow (ADAF) will occur at the time of $t_{\rm ADAF}$, when much of
the mass will be ejected prior to reaching the neutron star
surface. The X$-$ray luminosity of the neutron star will decline
rapidly with time due to $\dot{M}_{\rm X}$ (the rate of accretion
onto the star surface) diminishing. Therefore, the bright AXPs may
exist only between $t_{\rm trans}$ and 2$t_{\rm ADAF}$ (about 5000
$-$ 40000 years).

(d) Finally, accretion ceases since the matter of the disk is
exhausted. Thus no accretion$-$driven X$-$ray pulsar can be
observed and neither can radio emission be emitted owing to the
long rotational period of the neutron star (Qiao \& Zhang 1996,
Zhang et al. 2000). Consequently, the neutron star may be observed
as a dim isolated neutron star, a weak soft X$-$ray source powered
by rotational energy loss or thermal radiation.

\section{Conclusions and discussions}

The accretion disk model for periodic timing variations of pulsars
(e.g. PSR B1828$-$11) can elucidate the observed precession
period.
%
%avoiding the puzzle of the damping-out effect due to the vortices
%in the superfluid interior of the neutron star.
%
According to the analysis of the possible evolutionary scenario
discussed above, the PSR B1828$-$11 may be just in the ``radio
pulsar phase'' before the ``sleeping phase'' and the ``propeller
phase''. If our model proves to be true, the link between radio
pulsars and AXPs may be established and the fossil disk model for
AXPs may be strengthened. It also suggests that the neutron
star$-$fossil disk system might exist. Some discussions are listed
below.

1. Sarazin et al. (1980) pointed out when the angular momentum of
the disk is larger than that of the neutron star, the neutron star
will precess due to the effect of the accretion disk and the disk
will maintain its orientation. This is just our case for the
precession of the PSR B1828$-$11.

2. Why is the radio emission expected in our model but not
expected in the case of close X$-$ray binary system? Illarionov \&
Sunyaev (1975) suggested that the large optical depth of the
stellar wind for free$-$free absorption hinders the detection of
radio pulsars. In our model, a fossil disk is formed and there is
no enough stellar wind to absorb the radio emission. A case in
point is the PSR B1259$-$63, which has a $B_{\rm e}$ companion
star with high ellipticity and long orbital period. This neutron
star is observed as a radio pulsar when the companion encircles
far away from it. But no radio emission is detected when the
companion approaches very near the pulsar (Johnston et al. 1992).

3. The condition when the pulsar radio emission quenches is not
clear yet. A possible case may be that the cavity radius becomes
smaller than the light cylinder radius (Henriches 1983). In our
model, although the corotation radius of the PSR B1828$-$11
$R_{\rm co}=(GM/ \omega^{2})^{1/3}\approx9.19\times10^{7}$cm is
one order of magnitude smaller than its light cylinder radius
$R_{\rm LC}=c/ \omega=1.93\times10^{9}$cm, and the disk radius $R$
approaches $2-3R_{\rm co}$, this neutron star is still in the
``radio pulsar phase''. This presents an example for further
investigations.

4. We have ignored the angular momentum of the pinned superfluid
in the dynamic equations of the calculation. Can the pinning force
be strong enough to damp any precession of neutron stars? In
principle, detailed computations may determine whether the
vortices are pinned, but a definitive conclusion on the nature has
not been reached yet due to various uncertainties in the
microscopic physics.
Based on the calculation of Shaham (1977) and Sedrakian et al.
(1999), Stairs et al. (2000) conclude that the torque$-$free
precession of PSR B1828$-$11 should be damped out due to the
dissipation caused by the vortices pinning to the stellar crust on
the timescale of several hundred precession periods.
Nevertheless, a further consideration of Shaham (1986) indicates
an unpinned superfluid in the case of precession. It is also noted
by Link \& Cutler (2002), that hydrodynamic forces present in a
precessing star are probably sufficient to unpin all of the
vortices of the inner crust.
In addition, the 35$-$d periodicity of the accreting binary system
(Her X$-$1) has been interpreted as precession (Shakura, Postnov
\& Prokhorov 1998, and the references therein), which could be
another example of forced precessions.
We have therefore a tendency to suggest that, the observation of
PSR B1828$-$11 precession may imply that the vertex pinning in
this pulsar is much weaker than that predicted previously, at
least it is weaker than the possible disk$-$driven precession
effect.

5. In fact, the phenomenon of the secular periodic timing
variations, which also exists in many other radio pulsars (e.g.
Crab and Vela pulsar), has been investigated both observationally
and theoretically (Sedrakian et al. 1999 and the references
therein). If a neutron star were a rigid or semirigid solid star,
it would precess (Pines \& Shaham 1974); as is mentioned above,
the interaction between the superfluid vortices and the stellar
crust needs further considerations.
%
%but because of the
%interaction between the solid crust's nuclei and its internuclear
%superfluid neutrons' vortex lines, the damping time of the crustal
%precession should be several orders of magnitude smaller than a
%year (Shaham 1977, Sedrakian et al. 1999, Link \& Epstein 2000).
%
Ruderman (2001) pointed out that this is a crucial problem which
will raise doubts about canonical descriptions of neutron star
structure until it is solved.
%
%Even if the precession is due to forced precession,
No matter the precession is torque$-$free or forced, the neutron
star structure still needs to be investigated more. Instead of
neutron star, Xu (2003) proposed a solid strange star model, which
may be possible and also needs to be investigated more.

6. Our model cannot exclude the possibility of a companion star
(instead of a fossil disk) which can produce the precession
phenomenon as well. In this case, two limits should be considered:
the Roche Limit $d_{\rm R}$ (if the companion approached the
neutron star closer than this limit, it would be torn asunder by
tidal forces) and the Instability Limit $d_{\rm I}$ (if the
company strayed farther than this limit, it would escape due to
the differential perturbations from other celestial bodies). The
details of this case will be studied in another paper.

%______________________________________________________________

\begin{acknowledgements}

We are grateful to K. J. Lee et al. for their helpful discussions.
This work is supported by National Nature Science Foundation of
China (10173002, 10073001).

\end{acknowledgements}

%-------------------------------------------------------------

   \begin{figure}
   \centering
   \includegraphics[width=8.5cm]{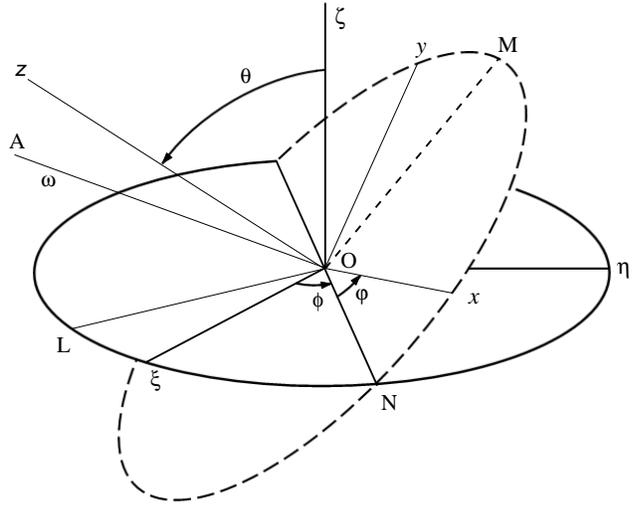}
      \caption{The geometry and parameters for
the calculations. In our calculation, the rotation axis $OA$ is
aligned with the principal axis $Oz$ of moment of inertia of the
neutron star.}
      \end{figure}
%
%_____________________________________________________________

%-------------------------------------------------------------
   \begin{figure}
   \centering
   \includegraphics[width=8.5cm]{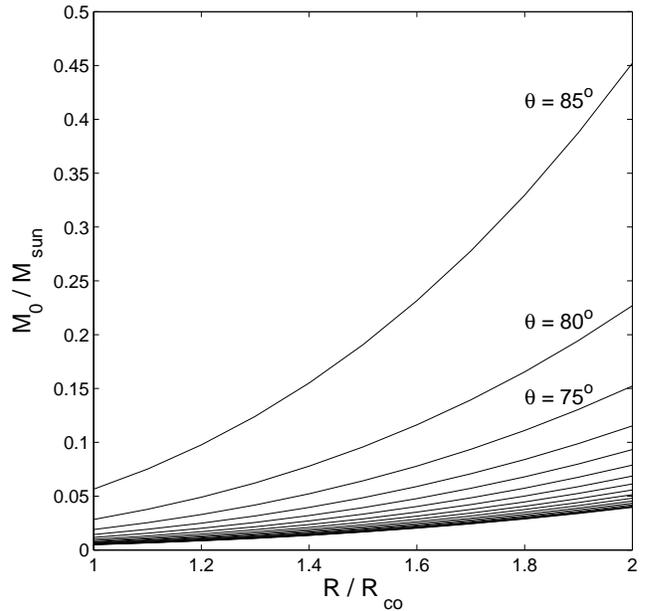}
      \caption{The relations between $M_{0}$ and
$R$ derived from the 1000$-$day precession period for a group of
$\theta$.}

   \end{figure}


\begin{thebibliography}{}

\bibitem[]{} Chatterjee, P., Hernquist, L., Narayan, R. 2000, ApJ,
534, 373

\bibitem[]{} Cordes, J. M. 1993, in ASP Conf. Ser. 36, Planets
around Pulsars, 43

\bibitem[]{} Henriches, H. F. 1983, in ``Accretion$-$driven stellar
X$-$ray sources'', eds: lewin, W. H. G., van den Heuvel E. P. J.,
393, Cambridge Univ. press

\bibitem[]{} Illarionov, A. F., Sunyaev, R. A. 1975, Astron. Ap.,
39, 185

\bibitem[]{} Johnston, S., Manchester, R. N., Lyne, A. G., Bailes,
M., Kaspi, V. M., Qiao, G. J., D'Amico, N. 1992, ApJ, 387, L37

\bibitem[]{} Jones, D. I., Andersson, N. 2001,
Mon.Not.R.Astron.Soc, 324, 811

\bibitem[]{} Link, B., Cutler, C. 2002, Mon.Not.R.Astron.Soc., 336, 211

\bibitem[]{} Link, B., Epstein, R. I. 2001, ApJ, 556, 392

\bibitem[]{} Nagase, F. 1989, Publ.Astron.Soc.Japan, 1

\bibitem[]{} Pines, D., Shaham, J. 1974, ComAp, 6, 37

\bibitem[]{} Qiao, G. J., Cheng, J.H. 1989, ApJ, 340, 503

\bibitem[]{} Qiao, G. J., Zhang, B. 1996, A\&A, 306, L5

\bibitem[]{} Rezania, V. 2002, preprint (astro$-$ph/0205180)

\bibitem[]{} Ruderman, M., 2001 (astro$-$ph/0109353)

\bibitem[]{} Sarazin, C. L., Begelman, M. C., Hatchett, S. P. 1980,
ApJ, 238, 129

\bibitem[]{} Sedrakian, A., Wasserman, I., Cordes, J. M. 1999, ApJ,
524, 34

\bibitem[]{} Shabanova, T. V., Lyne, A. G., Urama, J. O. 2001, ApJ,
552, 321

\bibitem[]{} Shaham, J. 1977, ApJ, 214, 251

\bibitem[]{} Shaham, J. 1986, ApJ, 310, 780

\bibitem[]{} Shakura, N. I., Postnov, K. A., Prokhorov, M. E.
1998, Astron.Astrophys., 331, L37

\bibitem[]{} Stairs, H., Lyne, A. G., Shemar, S. L. 2000, Nature,
406, 484

\bibitem[]{} Trumper, J. 1983, MitAG, 58, 7

\bibitem[]{} Xu, R. X. 2003, preprint (astro$-$ph/0302165)

\bibitem[]{} Xue, Y. Q., Qiao, G. J., Xu, R. X., Wang, H. G. 2003,
Acta Scientiarum Naturalium, Universitatis Pekinensis, accepted

\bibitem[]{} Zhang, B., Harding, A. K., Muslimov, A. 2000, ApJ,
531, L135

\end{thebibliography}
\end{document}